\documentclass[aps,prl,nobibnotes,twocolumn,superscriptaddress,shortbibliography]{revtex4-1}

\usepackage{amsfonts}
\usepackage{mathrsfs}
\usepackage{amsmath}
\usepackage{color}
\usepackage{graphicx}
\usepackage{bm}
\usepackage{amssymb}
\usepackage{xspace}
\usepackage{epstopdf}
\usepackage{dcolumn}
\usepackage{longtable}
\usepackage{multirow}
\usepackage{float}
\usepackage{comment}
\usepackage{lineno}

\providecommand{\tabularnewline}{\\}

\usepackage[colorlinks=true, letterpaper=true, pdfstartview=FitV,  linkcolor=blue, citecolor=blue, urlcolor=blue]{hyperref}

\makeatother

\begin{document}
\title{Single Pair of Weyl Points in Nonmagnetic  Crystals}
\author{Xiaotian Wang}
\affiliation{School of Physical Science and Technology, Southwest University, Chongqing 400715, China.}

\author{Feng Zhou}
\altaffiliation{X. Wang and F. Zhou contributed equally to this work.}
\affiliation{School of Physical Science and Technology, Southwest University, Chongqing 400715, China.}

\author{Zeying Zhang}
\affiliation{College of Mathematics and Physics, Beijing University of Chemical Technology, Beijing 100029, China}

\author{Weikang Wu}
\affiliation{Research Laboratory for Quantum Materials, Singapore University of
Technology and Design, Singapore 487372, Singapore}

\author{Zhi-Ming Yu}
\affiliation{Centre for Quantum Physics, Key Laboratory of Advanced Optoelectronic Quantum Architecture and Measurement (MOE), School of Physics, Beijing Institute of Technology, Beijing, 100081, China}
\affiliation{Beijing Key Lab of Nanophotonics \& Ultrafine Optoelectronic Systems, School of Physics, Beijing Institute of Technology, Beijing, 100081, China}

\author{Shengyuan A. Yang}
\affiliation{Research Laboratory for Quantum Materials, Singapore University of
Technology and Design, Singapore 487372, Singapore}

\begin{abstract}
Topological semimetal states having the minimal number, i.e., only a single pair, of Weyl points are desirable for the study of effects associated with chiral topological charges. So far, the search for such states is focused on magnetic spinful systems. Here, we find that nonmagnetic spinless systems can host a class of single-pair-Weyl-point (SP-WP) states, where the two Weyl points are located at two high-symmetry time-reversal-invariant momenta.  We identify 38 candidate space groups that host such states, and we show that the chiral charge of each Weyl point in the SP-WP state must be an even integer. Besides achieving the minimal number, Weyl points in SP-WP states are far separated in momentum space, making the physics of each individual point better exposed. The large separation combined with the even topological charge lead to extended surface Fermi loops with a non-contractible winding topology on the surface Brillouin zone torus, distinct from conventional Weyl semimetals. We confirm our proposal in the phonon spectra of two concrete materials TlBO$_2$ and KNiIO$_6$. Our finding applies to a wide range of systems, including electronic, phononic, and various artificial systems. It offers a new direction for the search of ideal platforms to study chiral particles.

\end{abstract}
\maketitle


{\emph{\textcolor{blue}{Introduction.}}}
The exploration of novel band degeneracies in crystals
and the associated physical properties is under rapid development \cite{Chiu2016Classification-RoMP,doi:10.1146/annurev-conmatphys-031113-133841,doi:10.1146/annurev-conmatphys-031016-025458,Armitage2018Weyl-RoMP,YU2022375}. One most  important type is the Weyl points (WPs) \cite{Wan2011Topological-PRB,Shuichi_Murakami_2007,PhysRevLett.107.127205,PhysRevLett.107.186806,Fang2012Multi-Prl}, which refer to a class of twofold degenerate nodal points carrying chiral topological charges. In a three-dimensional (3D) crystal, a WP acts as a point source of Berry curvature fields in momentum space, and its topological charge corresponds to the quantized Chern number ($\mathcal{C}$) from integrating the Berry flux on a small Gaussian sphere enclosing the WP. Among the WPs, the elementary one with unit charge was discovered first, and its existence does not require any symmetry (except for the lattice translations). Later, it was shown that WPs with $|\mathcal{C}|>1$ also exist with the help of crystalline symmetries \cite{PhysRevLett.107.186806,Fang2012Multi-Prl}. Recent studies demonstrate that the maximum $\mathcal{C}$ of a single WP can be up to four \cite{YU2022375,PhysRevB.102.125148,FHH-npj,LiuGB-PRB2022,Zhangzy-2022}. The nontrivial chiral charges lead to fascinating consequences, such as Fermi arc surface states \cite{Wan2011Topological-PRB,PhysRevX.5.031013,yang2015weyl,doi:10.1126/science.aaa9297}, chiral/axial anomaly \cite{NIELSEN1983389,PhysRevB.86.115133,PhysRevB.88.104412,PhysRevLett.111.027201,PhysRevB.87.235306,PhysRevX.4.031035,PhysRevX.5.031023,Burkov_2015}, chiral Landau bands \cite{NIELSEN1983389,PhysRevLett.111.246603,PhysRevLett.126.046401,doi:10.1126/science.aac6089}, and etc.

{
\global\long\def\arraystretch{1.4}%

\begin{table*}
\caption{\label{Tab} The candidate space groups that can host SP-WP states.
HSP denotes the high-symmetry point, where the WPs are located. PG denotes the point group at the HSP. IRR stands for the irreducible (co-)representation of the little group associated to the WPs.
For SG 75, 77, 89 and 93, the two WPs  can locate at any two of the four TRIM points. }
\begin{ruledtabular}
\begin{tabular}{lllll}
Species & SGs (HSPs) & PG  & IRR & Example\tabularnewline
\hline
Charge-2 WP &  75 and 77 ($\Gamma$, M, Z, A); 76 ($\Gamma$, M); 78 ($\Gamma$, M); 79 and 80 ($\Gamma$, Z) & $C_{4}$  & $\{R_{2},R_{4}\}$ & TlBO$_{2}$\tabularnewline
 & 89 and 93 ($\Gamma$, M, Z, A); 90, 94, 97 and 98  ($\Gamma$, Z);  91 and 95 ($\Gamma$, M) & $D_{4}$  & $R_{5}$ & \tabularnewline
 & 143-145 ($\Gamma$, A); 146 ($\Gamma$, Z) & $C_{3}$  & $\{R_{2},R_{3}\}$ & B$_3$N$_3$Cl$_6$\tabularnewline
 & 149-154 ($\Gamma$, A); 155 ($\Gamma$, Z) & $D_{3}$  & $R_{3}$ & KNiIO$_{6}$\tabularnewline
 & 168 ($\Gamma$, A); 171-172 ($\Gamma$, A) & $C_{6}$  & $\{R_{2},R_{6}\}$ or $\{R_{3},R_{5}\}$ & \tabularnewline
 & 177 ($\Gamma$, A); 180-181 ($\Gamma$, A) & $D_{6}$  & $R_{5}$ or $R_{6}$ & \tabularnewline
\hline
Charge-4 WP & 195 ($\Gamma$, R); 197 ($\Gamma$, H) & $T$  & $\{R_{2},R_{3}\}$  & \tabularnewline
 & 199 ($\Gamma$, H) & $T$  & $\Gamma$: $\{R_{2},R_{3}\}$, H: $\{R_{5},R_{6}\}$ & \tabularnewline
 & 207 and 208 ($\Gamma$, R); 211 ($\Gamma$, H) & $O$  & $R_{3}$ & \tabularnewline
 & 214 ($\Gamma$, H) & $O$  & $\Gamma$: $R_{3}$, H: $R_{6}$ & \tabularnewline

\end{tabular}\end{ruledtabular}
\end{table*}

}
Currently, a major challenge in the field is to find an ideal Weyl semimetal state hosting the WPs \cite{Armitage2018Weyl-RoMP,doi:10.1146/annurev-conmatphys-031016-025458}. Here, a crucial condition for being ``ideal'' is that the system should have a \emph{minimal} number of WPs. The Nielsen-Ninomiya no-go theorem \cite{NIELSEN198120,NIELSEN1981173} dictates that a net chiral charge cannot exist alone in the Brillouin zone (BZ), so there should be at least one pair of WPs with opposite charges~\footnote{We consider the Weyl semimetal state, where all coexisting band degeneracies are WPs. This excludes the case when a WP coexists with charged nodal surfaces, as in Ref. \cite{Yu2019Circumventing-PRB}}. Despite much effort in
searching for WPs in real materials and artificial systems \cite{PhysRevB.89.081106,PhysRevX.5.031013,WengPRX_WP_2015,PhysRevX.5.031023,soluyanov2015type,PhysRevLett.114.206602,PhysRevB.92.161107,doi:10.1126/science.aaa9273,ruan2016symmetry,PhysRevLett.117.236401,
doi:10.1126/sciadv.1600295,PhysRevLett.116.226801,belopolski2017signatures,doi:10.1126/science.aaq1221,doi:10.1126/science.aaq1221,Zhang2018Double-Prl,XuH_prl_2019,PhysRevLett.124.105303,li2021computation,xiao2015synthetic,
PhysRevLett.117.224301,doi:10.1126/science.aav2873,
PhysRevLett.122.104302,10.1093/nsr/nwaa192,PhysRevLett.125.143001}, so far, the minimal number of WPs, i.e., a single pair, has only been achieved in very few magnetic systems, such as electronic WPs in magnets MnBi$_2$Te$_4$  \cite{doi:10.1126/sciadv.aaw5685,PhysRevLett.122.206401} and EuCd$_2$As$_2$ \cite{PhysRevB.99.245147,PhysRevB.100.201102,doi:10.1126/sciadv.aaw4718} under negative pressure or external fields, and magnonic WPs in some magnetic pyrochlores \cite{PhysRevLett.117.157204}, but not in nonmagnetic systems. In fact, it was widely believed that in nonmagnetic systems, there should be a minimal of four WPs \cite{Wan2011Topological-PRB,doi:10.1146/annurev-conmatphys-031113-133841,Armitage2018Weyl-RoMP}. The argument goes as follows. Suppose a WP with charge $\mathcal{C}$ is located at point $\bm k$ in the BZ. If the time reversal symmetry $\mathcal{T}$ is preserved, there must be another WP at $-\bm k$ with the same charge $\mathcal{C}$. However, according to the no-go theorem \cite{NIELSEN198120,NIELSEN1981173}, these two WPs cannot exist alone: there should be at least another two WPs (each with charge $-\mathcal{C}$) to ensure the chiral charge neutrality of the BZ. Because of this, the search for a single pair of WPs is focused on magnetic systems with broken $\mathcal{T}$.

Importantly, there is a flaw in the above argument: if the first WP is located at a time-reversal-invariant momentum (TRIM), then $\bm k$ and $-\bm k$ refer to the same point, and $\mathcal{T}$ would not lead to the WP doubling. Nevertheless, here, we need to distinguish spinful and spinless systems. Particularly, a single pair cannot be achieved in spinful systems: the algebra $\mathcal{T}^2=-1$ dictates Kramers degeneracy at all eight TRIM points in BZ, such that the resulting WPs (known as Kramers WPs \cite{ChangNM2018}) have a minimal number of eight. In contrast, for spinless systems, $\mathcal{T}^2=1$, so they do not have this constraint.

In this work, we demonstrate that it is indeed possible to realize a single pair of WPs in nonmagnetic spinless systems. We analyze the symmetry requirements to achieve such single pair of WPs state (denoted as SP-WP state) and develop a strategy to search for them. We find that SP-WP state can be hosted in 38 {{(out of 230)}} space groups (SGs), and we prove that the topological charge of each WP here must be even. We then demonstrate our idea by
identifying realistic material candidates, including TlBO$_2$ and  KNiIO$_6$. In both materials, a single pair of WPs (with $\mathcal{C}=\pm 2$)
appear in the phonon spectrum, as the only band degeneracies in a frequency range. We note that besides achieving the minimal number of WPs, these systems have another advantage, namely, the two WPs are far separated in BZ, such that the physics of each individual WP can be better exposed. Furthermore, the large separation combined with the even chiral charges lead to extensive surface Fermi loops forming non-contractible winding topology on the BZ torus. Our proposal opens a new direction for realizing ideal Weyl states. It applies to systems ranging from electronic, phononic, to various artificial systems, and will facilitate the study of fascinating physics associated with chiral topological charge.

{\emph{\textcolor{blue}{General analysis.}}}
We begin by analyzing the symmetry conditions for our proposed SP-WP state. Assume the two WPs are residing at points $\bm k_1$ and $\bm k_2$, respectively. The following conditions should be satisfied.

(i) First, as discussed, $\bm k_1$ and $\bm k_2$ must be TRIM points. Moreover, they must form a closed set under all symmetry operations of the system, because any symmetry operation that transforms $\bm k_1$ or $\bm k_2$ to another point $\bm k_3$ must generate a third WP.

(ii) The crystal symmetry operations are divided into two classes: proper and improper. A proper (improper) operation preserves (flips) the orientation of space, so it keeps (reverses) the chiral charge $\mathcal{C}$ of a WP. Combined with (i), this means $\bm k_i$ ($i=1,2$) is invariant under proper operations, whereas $\bm k_1$ and $\bm k_2$ are interchanged under any improper operation. Namely,
\begin{equation}
  \mathcal{O}\bm k_i=\bm k_i, \qquad i=1,2
\end{equation}
if the symmetry operation $\mathcal{O}$ is proper, and
\begin{equation}
  \mathcal{O}\bm k_1=\bm k_2,\qquad \mathcal{O}\bm k_2=\bm k_1,
\end{equation}
if $\mathcal{O}$ is improper. Clearly, the inversion symmetry must be broken, since all TRIM points are also inversion-invariant points.

(iii) Since $\bm k_i$ $(i=1,2)$ is a TRIM point, the little group $M^{\boldsymbol{k}_{i}}$  there can be generally expressed as
\begin{equation}
M^{\boldsymbol{k}_{i}}  = G^{\boldsymbol{k}_{i}}+{\cal{T}} G^{\boldsymbol{k}_{i}},
\end{equation}
where $G^{\boldsymbol{k}_{i}}$ is the corresponding crystallographic little group and $\mathcal{T}^2=1$. To realize a WP at $\bm k_i$, a necessary condition is that $M^{\boldsymbol{k}_{i}}$
must have at least one single-valued irreducible co-representation with dimension 2.

Based on these conditions, we search through all the 230 SGs for nonmagnetic systems to screen out the candidates that may host our proposed SP-WP state. Specifically, for each SG, we check all its eight TRIM points in BZ, looking for those satisfying the conditions (i)-(iii). Then, for the resulting ones, we construct the $k\cdot p$ effective model corresponding to each 2D irreducible co-representations in (iii), to verify whether it is indeed a WP and (if so) to obtain its topological charge $\mathcal{C}$. The results are presented in Table~\ref{Tab},
in which we list the candidate SGs, the possible locations of the WPs, and the corresponding symmetry representation.
This offers guidance to the search and design of systems hosting a single pair of WPs.

We comment that our discussion applies to a wide range of physical systems, including electrons in materials with weak spin-orbit coupling, phonons in real materials, and various classical waves in artificial systems (e.g., acoustic/photonic crystals, electric circuit arrays, mechanical networks). In this work, we shall take phonons in two real materials as examples.

{\emph{\textcolor{blue}{Even topological charge.}}}
Interestingly, we find that for a spinless nonmagnetic crystal, the topological charge $\mathcal{C}$ of a WP located at a TRIM point must be even.
This can be proved by contradiction. Assume there is a WP with odd $\mathcal{C}$ at a TRIM point $\bm K$. To be specific, let's first consider $\mathcal{C}=1$. This means the Chern number is unity for the 2D insulating spectrum defined on a small sphere $S^2$ centered at $\bm K$.
Then, we exert a perturbation on the system (e.g., lattice strain) to break all the crystallographic symmetries of $G^{\bm K}$ (which is always possible). Under the perturbation, the WP remains existing due to its topological stability,
but its position will deviate from the TRIM point $\bm K$, as $\mathcal{T}$ alone (for a spinless system) cannot pin any WP there. For weak enough perturbation, the WP remains within the sphere $S^2$, and $\mathcal{C}$ on $S^2$ should not change. Let's denote the WP's position by $\bm K+\bm q$, with $\bm q$ the small deviation. Then, under $\mathcal{T}$, there must be an extra $\mathcal{C}=1$ WP at $\bm K-\bm q$. This means that the $\mathcal{C}$ on $S^2$ must change its parity (jump from one to an even integer) under an arbitrarily weak perturbation, which is impossible~\footnote{The other WP of the pair is at another TRIM point, which is far away from $\bm K$, so it does not affect the local analysis.}. The argument can be readily extended to any WP with an odd $\mathcal{C}$ larger than one.

Note that the maximal chiral charge for a WP stabilized in a crystal is four. Thus, according to the argument above, the charge for the WPs in the SP-WP state can only be two or four. This is confirmed by the results in Table~\ref{Tab}.

\begin{figure}[b]
\includegraphics[width=8cm]{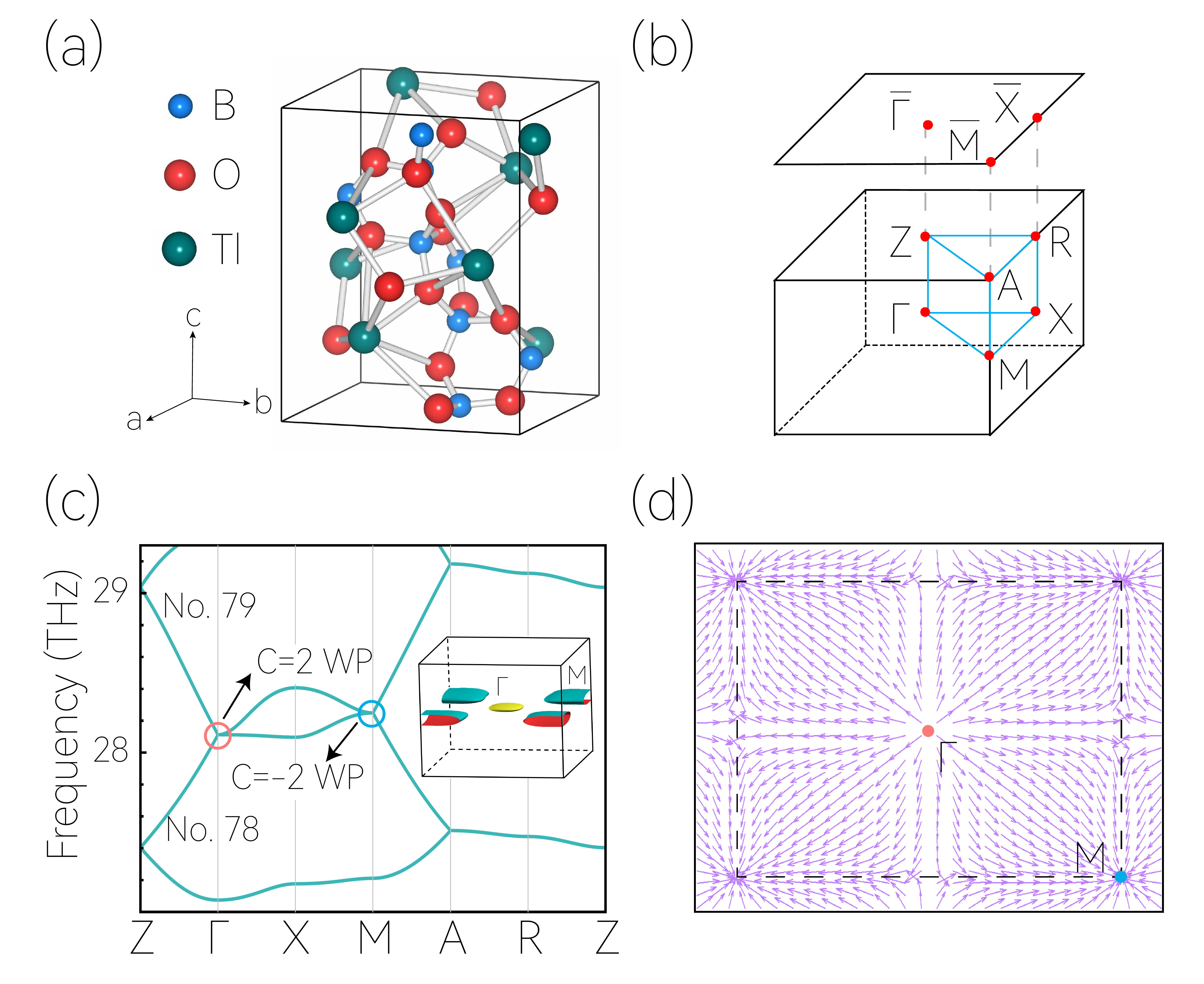}
\caption { (a) Crystal structure of TlBO$_2$. (b) Bulk and surface BZs of TlBO$_2$. (c) Calculated phonon spectrum of TlBO$_2$ along high-symmetry paths. The inset shows the constant energy contour at $28.22$ THz. (d) Distribution of Berry curvature in the $k_z$ = 0 plane, which contains the two WPs.
\label{fig1}}
\end{figure}

The even WP charge determines that there are an even number of Fermi arcs emanating from the projection of the WP on a surface. Combined with two characters of the SP-WP state, namely, the two WPs being at TRIM points and $\mathcal{T}$ being preserved, it leads to an distinct feature of topological surface modes: The surface zero-modes would form \emph{closed} and \emph{non-contractible} Fermi loops with a nontrivial winding topology on the surface BZ torus. This is in contrast to conventional Weyl semimetals, where the surface modes are of the typical form of \emph{open} Fermi arcs~\cite{Wan2011Topological-PRB}. We shall elaborate more on this point when discussing concrete examples below.

{\emph{\textcolor{blue}{Example 1: TlBO$_2$}.}}
The first example is phonons in a concrete material TlBO$_2$. Phonons are naturally spinless particles. The crystal TlBO$_2$ preserves $\mathcal{T}$ and belongs to SG~76 (P4$_1$)~\cite{TOUBOUL197739}, one of the candidates in Table~\ref{Tab}. Its crystal structure is illustrated in Fig.~\ref{fig1}(a).

\begin{figure}[b]
\includegraphics[width=8cm]{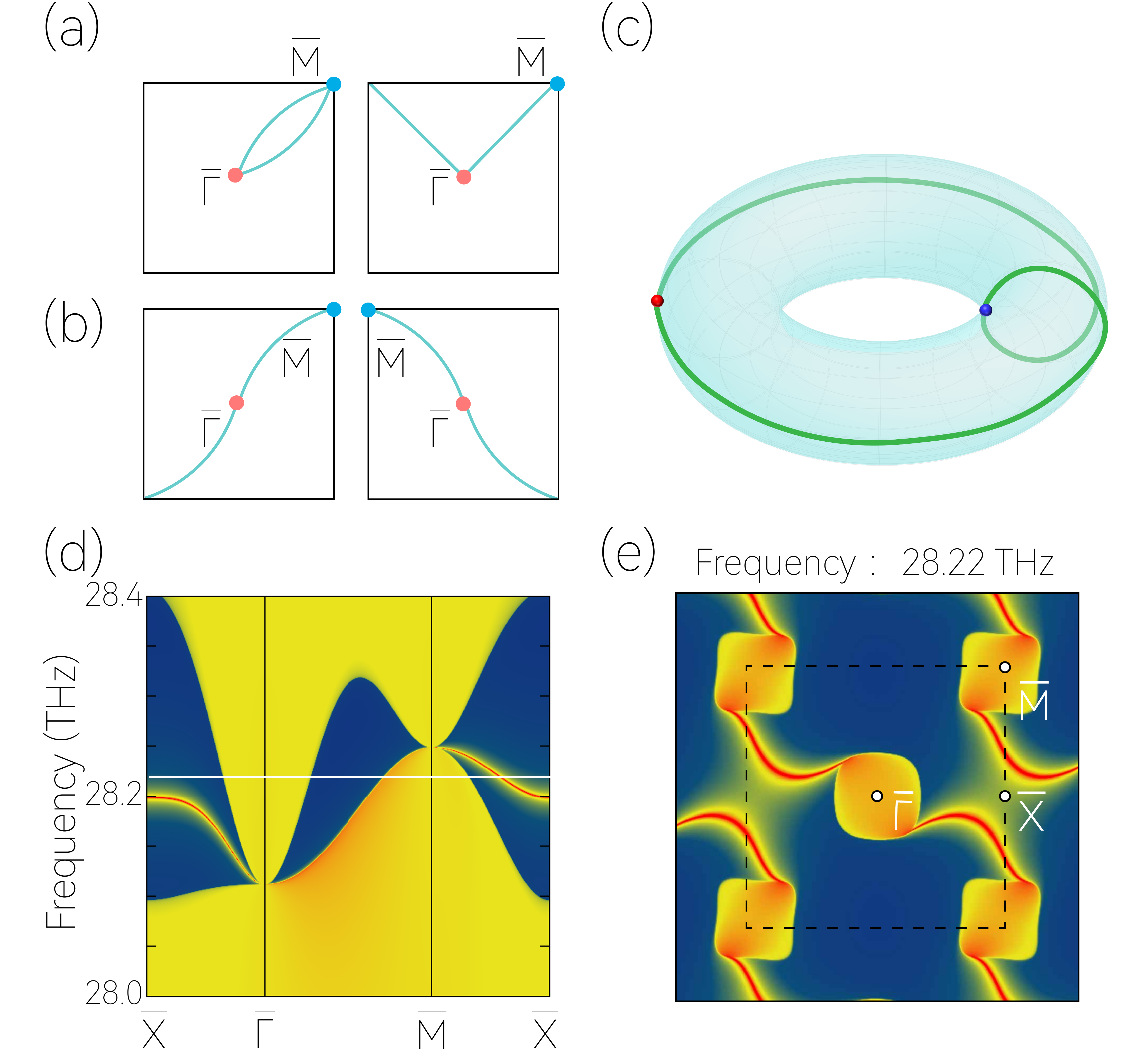}
\caption{ (a) Impossible and (b) possible pattern of surface zero-modes on the (001) surface of TlBO$_2$.
Red and blue points are the surface projections of the two bulk WPs.
(c) shows schematically  the non-contractible winding of the Fermi loop on the surface BZ torus.
(d) Projected spectrum on the (001) surface of TlBO$_2$. (e) shows the  constant energy slice at $28.22$ THz,  as marked by the white dashed line in (d).
\label{fig2}}
\end{figure}

According to Table~\ref{Tab},  SG~76 may host a single pair of WPs at $\Gamma$ and $M$ points. In
Fig.~\ref{fig1}(c), we plot the phonon band structure for TlBO$_2$ around 28 THz (the whole spectrum can be found in Supplemental Material (SM)~\cite{SM}). One observes that there are indeed two WPs at the proposed TRIM points, formed by the crossing of two phonon branches No.~78 and No.~79, which are the \emph{only} band degeneracies in this energy interval [see the inset of Fig.~\ref{fig1}(c)]. Around each WP, the band splitting is linear along $k_z$ ($c$ axis) and quadratic in the $k_x$-$k_y$ plane, consistent with its charge-2 (C-2) character. The derived $k\cdot p$ effective models for the two WPs are presented in SM ~\cite{SM}. We have verified that the WP at $\Gamma$ ($M$) has $\mathcal{C}=+2$ ($-2$). This can also be visualized in the plot of Berry curvature field in Fig.~\ref{fig1}(d). One can see that the field is emitted from the WP (positive charge) at $\Gamma$ and absorbed by the WP (negative charge) at $M$.

\begin{figure}
\includegraphics[width=8cm]{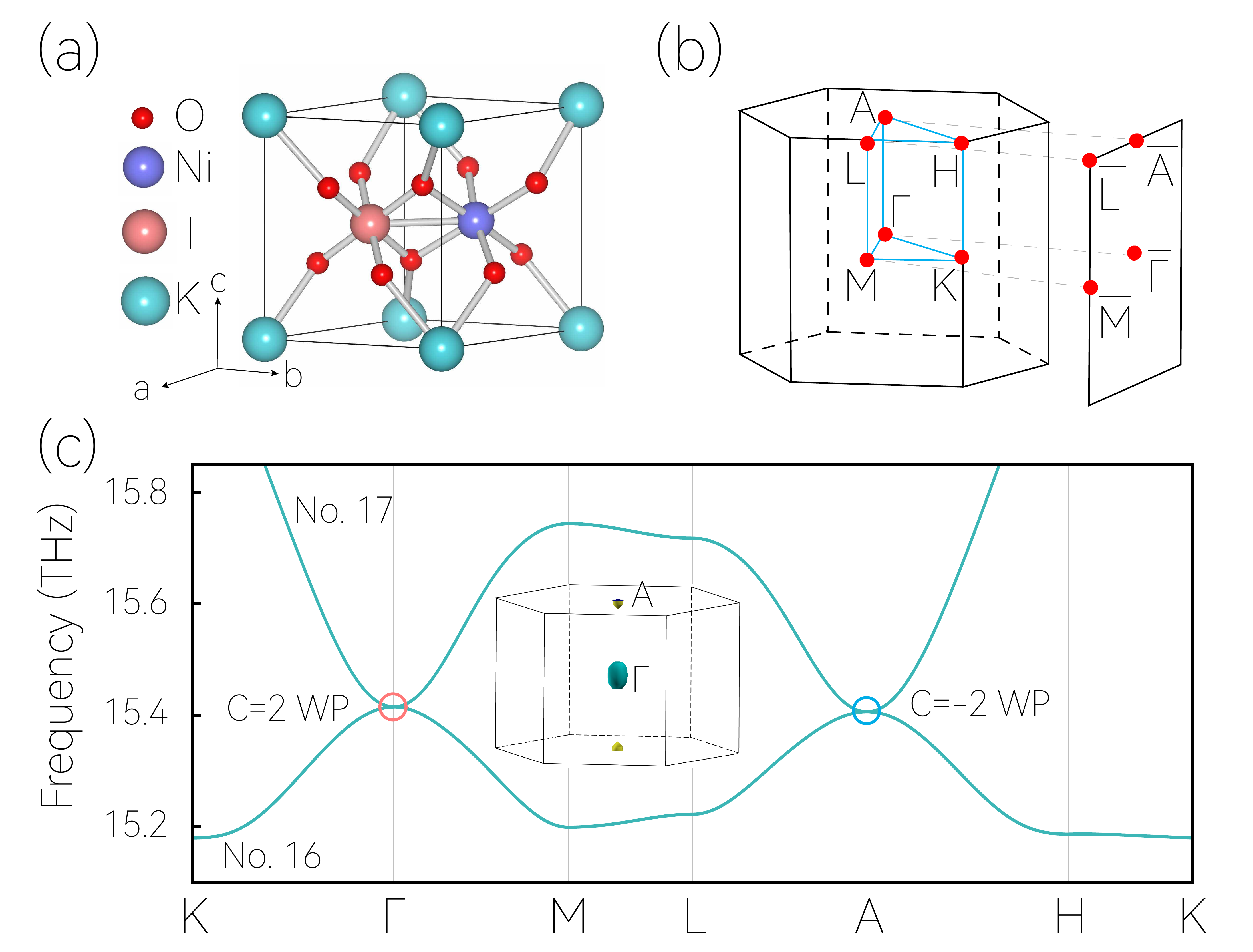}
\caption{ (a) Crystal structure of KNiIO$_6$. (b) Bulk and  surface BZs. (c) Calculated phonon spectrum of KNiIO$_6$ along high-symmetry paths. The inset shows the constant energy contour at $15.41$ THz.
\label{fig3}}
\end{figure}

Next, we investigate the surface modes associated with the SP-WP state in TlBO$_2$. Let's consider the $(001)$ surface.
The surface projections of the two bulk WPs are at $\bar{\Gamma}$
and $\bar{M}$, which are also TRIM points of the surface BZ. Since each WP has a chiral charge of two, there must be two Fermi arcs connected to its surface projection. Moreover, because the projected WPs are sitting at TRIM points, the two arcs must be time-reversal partners, i.e., under $\mathcal{T}$ operation, one arc is mapped to the other. It follows that the surface zero-modes may form the patterns in {{Fig.~\ref{fig2}(b) but not in Fig.~\ref{fig2}(a).}}
This point is confirmed by our calculation result in Fig.~\ref{fig2}(d-e), which show the projected spectrum for TlBO$_2$ on the $(001)$ surface.

Interestingly, the patterns in Fig.~\ref{fig2}(b) and Fig.~\ref{fig2}(e) represent a non-contractible closed loop with a nontrivial winding around BZ. To understand this, note that the surface BZ is a two-torus $T^2=S^1\times S^1$. A closed loop on $T^2$ can be characterized by the number of times it winds around each circle $S^1$ (corresponding to the fundamental homotopy group~\cite{PhysRevB.96.081106}). The surface Fermi loop in Fig.~\ref{fig2}(e) winds around each $S^1$ (i.e., each direction of the 2D BZ) once, which is schematically illustrated in Fig.~\ref{fig2}(c). Obviously, this winding pattern is topological, in the sense that it is stable against perturbations.
This type of surface Fermi loops is distinct from the open Fermi arcs for conventional Weyl semimetals~\cite{Wan2011Topological-PRB}. Our analysis shows that it is an important feature of SP-WP state.

\begin{figure}[t]
\includegraphics[width=8cm]{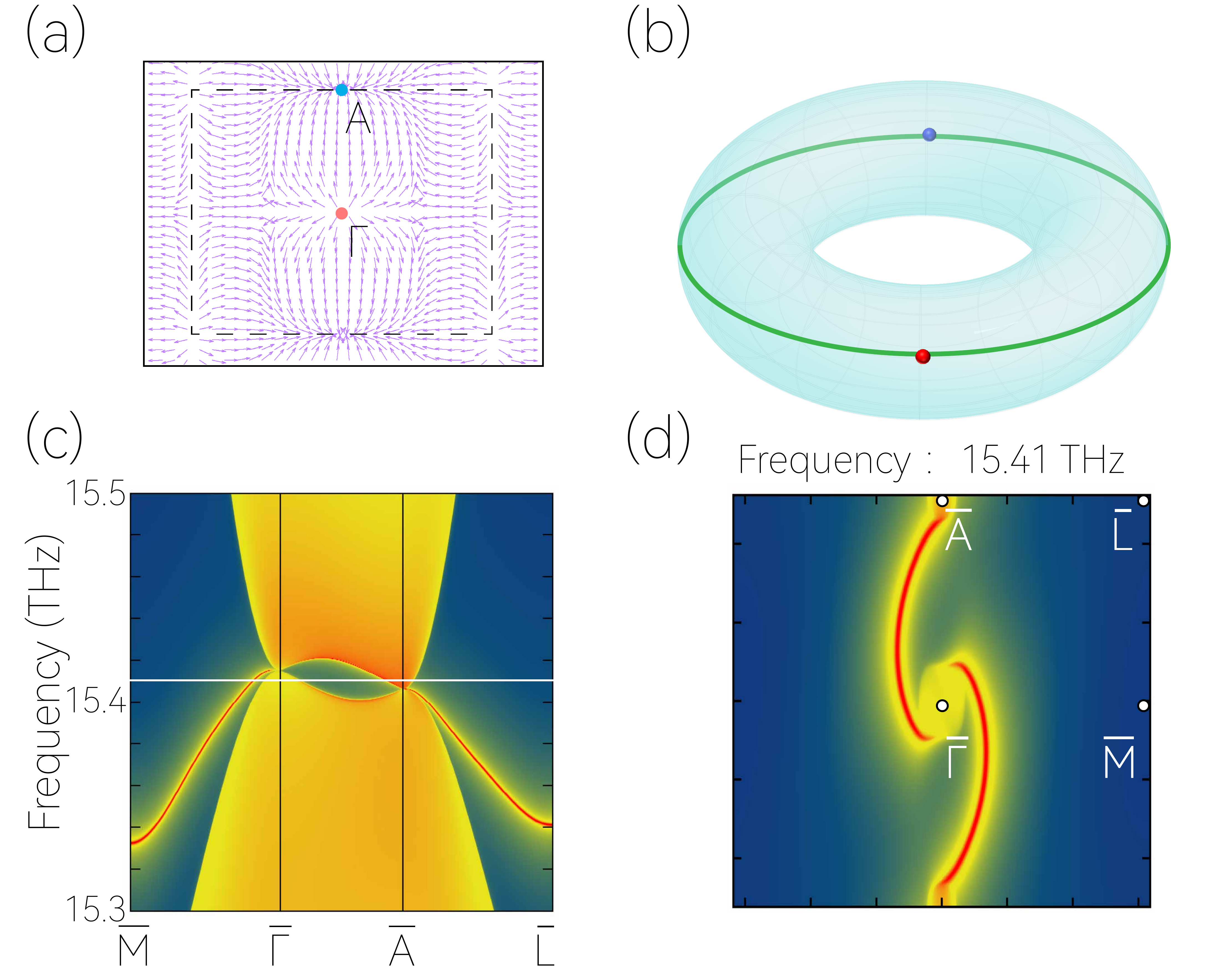}
\caption{(a) Distribution of Berry curvature in the $k_z=0$ plane. The dashed line indicates the first BZ. (b) shows schematically the winding pattern of surface Fermi loop for KNiIO$_6$'s side surface.
(c) Projected spectrum on the $(10\bar{1}0)$ surface. (d) shows the constant energy slice at $15.41$ THz, as marked by the white dashed line in (c).
\label{fig4}}
\end{figure}

{\emph{\textcolor{blue}{Example 2: KNiIO$_6$}.}}
Our second example is the phonons in KNiIO$_6$. The material belongs to SG~149 (P3$_1$2)~\cite{10.1139/v69-630}, and its lattice structure is illustrated in Fig.~\ref{fig3}(a).

In Fig.~\ref{fig3}(c), we plot its phonon band structure around $15.4$ THz, where a single pair of WPs are found at $\Gamma$ and $A$ points, consistent with Table~\ref{Tab}. Scanning through the BZ confirms that these are the only degeneracies in this energy interval. Our calculation shows that the WP at $\Gamma$ ($A$) carries $\mathcal{C}=+2$ $(-2)$. The corresponding Berry curvature field distribution is shown in Fig.~\ref{fig4}(a).

We find that the key features of the SP-WP state here can be captured by a simple two-band lattice model defined on a hexagonal lattice \cite{SM}
\begin{eqnarray}\label{ham}
H & = & t_1 \sin \frac{\sqrt{3} k_x}{2} \sin \frac{k_y}{2} \sigma_x \nonumber \\
 && + \left[ t_2 (2 \cos\frac{\sqrt{3} k_x}{2} \sin \frac{k_y}{2}-\sin k_y)+ t_3 \sin k_z \right]\sigma_y \nonumber \\
 &&-\frac{t_1}{\sqrt{3}} (\cos \frac{\sqrt{3} k_x}{2} \cos \frac{k_y}{2} -\cos k_y ) \sigma_z,
\end{eqnarray}
where the $t$'s are real hopping parameters, and $\sigma$'s are Pauli matrices corresponding to an orbital degree of freedom. This simple model may serve as a starting point for investigating the SP-WP state.

In Fig.~\ref{fig4}(c-d), we plot the projected spectrum on the $(10\bar{1}0)$ surface of KNiIO$_6$. Here, because the projected WPs are at $\bar{\Gamma}$ and $\bar{A}$ points, similar argument shows that surface zero-modes must form a closed non-contractible Fermi loop that winds around the surface BZ along only one direction, as illustrated in Fig.~\ref{fig4}(b), which has a different topology from that in Fig.~\ref{fig2}(c).

{\emph{\textcolor{blue}{Discussion}.}}
We have proposed the SP-WP state, a state with the minimal number of WPs, in nonmagnetic systems. As discussed, such spinless systems are ubiquitous. For example, besides phonons, materials with negligible spin-orbit coupling (like carbon allotropes \cite{PhysRevB.92.045108,Chen2015Nanostructured-Nl,Zhong2017Three-Nc}) are promising to realize SP-WP state in the electronic band structure. Moreover, artificial systems  such as acoustic crystals \cite{PhysRevLett.114.114301,ma2019topological,yang2019topological,PhysRevLett.122.136802,he2020observation,luo2021observation} offer a highly flexible platform to realize various spinless state. We expect our proposed SP-WP state can be readily achieved in such systems.

Compared to conventional Weyl states, the SP-WP state has several distinctions. The minimal number of WPs would simplify the theoretical analysis and interpretation of experimental results. The large separation between the two WPs is beneficial for exposing the physics of WPs without their intervening each other. In addition, the extended surface Fermi loops can be more easily imaged in  spectroscopy experiments \cite{doi:10.1021/acsnano.5b06807,zheng2018quasiparticle,RevModPhys.93.025002}.

\bibliography{SWP-ref}

\end{document}